\documentclass[twocolumn,showpacs,preprintnumbers,amsmath,amssymb]{revtex4}

\usepackage{graphicx}
\usepackage{dcolumn}
\usepackage{bm}

\begin{document}


\title{Analysis of the atom-number correlation function in a few-atom trap}

\author{Youngwoon \surname{Choi}}
\author{Seokchan \surname{Yoon}}
\author{Sungsam \surname{Kang}}
\author{Woongnae \surname{Kim}}
\author{Jai-Hyung \surname{Lee}}
\author{Kyungwon \surname{An}}
\email{Kwan@phya.snu.ac.kr}
\thanks{Fax: +82-2-884-3002}
\affiliation{School of Physics, Seoul National University, Seoul,
151-742, Korea}
\date{\today}

\begin{abstract}
Stochastic properties of loading and loss mechanism in a few atom trap are analyzed. An approximate formula is derived for the atom-number correlation function for the trapped atoms in the limit of reasonably small two-atom loss rate. Validity of the approximate formula is confirmed by numerical simulations.
\end{abstract}

\maketitle
\section{Introduction}
Techniques for trapping small number of atoms in a microscopic volume have recently become an important tool for wide range of experiments in atomic physics and quantum optics such as cold collisions \cite{cold-collision}, atom metrology \cite{metrology}, cavity quantum electrodynamics \cite{cavity-QED} and quantum information \cite{quantum-information}. 

There have been numerous measurements of trap loading and loss parameters of a magneto-optical trap (MOT) with many atoms, but they used indirect methods such as fitting to a model curve, $dn/dt=R-\Gamma n-\int\beta n^{2}$ \cite{beta-paper}, where $n$ is the density of trapped atoms, $\Gamma$ the loss rate due to collisions with background atoms, and $\beta$ a coefficient for two-atom collisional process among trapped atoms. Recently, several groups have trapped a small number of atoms in a MOT with a strong magnetic field gradient and could observe individual loading and loss events in real time. In this way, loading rate $R$, one-atom loss rate $L_{1}$, and two-atom loss rate $L_{2}$ have been directly measured \cite{two-atom-loss,yoon06}. 

The atom-number correlation function has also been measured from the observed sequence of instantaneous atom number in a trap \cite{correlation-measurement}. Since no analytic solution is known for the master equation in the limit of non-negligible two-atom loss, the study of the atom-number correlation function has been limited to the case of no two-atom loss, for which the correlation function does not provide any further information other than the one-atom loss rate. 

In the present work, we provide a comprehensive frame work for investigation of the atom-number correlation function based on the master equation (Chaps.\ II and III). An approximate formula has been derived for the correlation function in the limit of non-negligible two-atom loss (Chap. IV) and the validity of the approximate formula has been put to a test by numerical simulations (Chap.\ V).

\section{Description of Model}

The time dynamics of the atom number $N$ in a trap is governed by loading and loss processes. 
The loading process occurs at a certain rate $R$, called loading rate, which is determined by capture ability of the trap. The capture ability depends on such experimental conditions as laser intensity, laser beam size, laser-atom detuning and background source density of atoms but not on the number of atoms trapped already. We can assume $R$ as a constant under fixed experimental condition.

The loss rates, on the other hand, are affected by the number of atoms in the trap. One-atom loss occurs when one of the trapped atoms collides with a fast-moving background atom of different kind. One atom loss rate is thus linearly proportional to the number of atoms $N$ already present in the trap. We define $\Gamma_1$ as the one-atom loss {\em coefficient} in such a way that the one atom loss {\em rate} $L_1$ is given by $L_1=\Gamma_1 N$.
 
Two-atom loss process is due to the collision between two of the trapped atoms. 
There are several types of two-atom collision processes responsible for two-atom losses. They are ground-state hyperfine-changing collisions, ground-excited-state fine-structure changing collision and ground-excited-state radiative escape. Whenever these collisions occur, the colliding two atoms can gain enough kinetic energy for escaping from the trap \cite{two-atom-loss}. Therefore, the two-atom loss
rate is proportional to the number of two-atom combinations for the atoms in the trap. In terms of $\Gamma_2$  the two-atom loss {\em coefficient}, the two-atom loss {\em rate} $L_2$ is then given by $L_2=\Gamma_2 N(N-1)/2$. 

\begin{figure}
\includegraphics[width=3.4in]{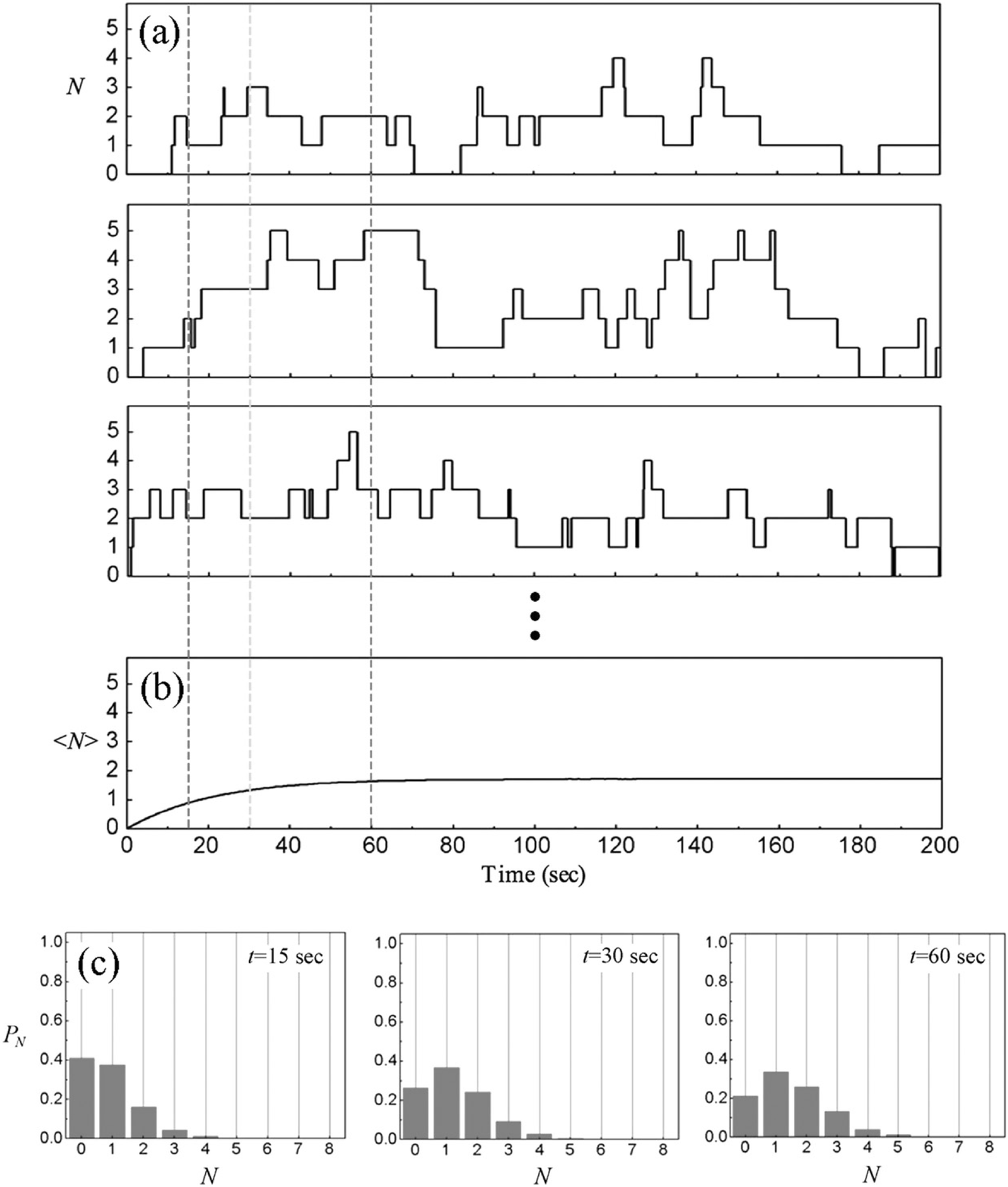}
\caption{(a) Ensemble picture for the loading process turned on at $t=0$. For each member of the ensemble, the atom number constantly fluctuates due to random loading and loss processes. (b) Ensemble averaged atom number in time. (c) Ensemble averaged atom number distribution at various times. }
\label{fig1}
\end{figure}

The loading and loss processes occur randomly. Due to random nature of these processes, it is more convenient to treat the problem in terms of the atom number distribution function $P_N$ than to deal with the time variation of the instantaneous atom number $N$ itself. In order to understand the connection between time evolution of $P_N(t)$ and that of $N$, let us suppose that we turn on the loading process at time $t=0$ and observe the atom number, initially zero, afterwards. The atom number will change randomly as depicted in Fig.\ \ref{fig1}, 
but the center of fluctuation will increase to a steady-state value determined by the balance between the loading and loss processes.

If we repeat the observation infinitely many times, each observation will give a different time sequence in detail. However, if we average all of the observed time sequences with the initial starting time aligned, we obtain a distribution of atom numbers at any time $t$, corresponding to $P_N(t)$, and a sequence of averaged atom number as a function of time. Alternatively, we can replace the infinitely many observations with an ensemble of identically prepared traps. The sequence of averaged atom number is then an ensemble average of the atom number $\langle N\rangle$.

\begin{figure}
\includegraphics[width=3in]{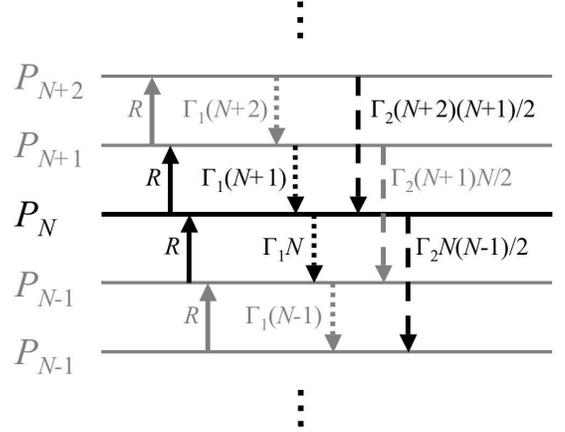}
\caption{Connections among probabilities $P_N$'s. Loading process $N-1\rightarrow N$, one-atom loss process $N+1\rightarrow N$ and two-atom loss process $N+2\rightarrow N$ result in an increase in $P_N$ whereas loading process $N \rightarrow N+1$, one-atom loss process $N\rightarrow N-1$ and two-atom loss process $N\rightarrow N-2$ cause a reduction in $P_N$.}
\label{P_N-connection}
\end{figure}

It is shown that the time evolution of the ensemble-averaged distribution $P_N(t)$ can be described the following master equation:
\begin{eqnarray}
\frac{dP_{N}}{dt}&=&RP_{N-1}+\Gamma_{1}(N+1)P_{N+1}\nonumber\\
& &+\frac{1}{2}\Gamma_{2}(N+2)(N+1)P_{N+2}\nonumber\\& & -\{R+\Gamma_{1}N+\frac{1}{2}N(N-1)\Gamma_{2}\}P_{N}\;.
\label{eq:master}
\end{eqnarray}
for $N=0,1,2,\ldots$ with a convention of $P_{-1}=0$. Connections among probabilities $P_N$'s are depicted in Fig.\ \ref{P_N-connection}.

\section{Without Two-atom Loss Terms}
If a density of atoms in a trap is low enough, the collisions among the trapped atoms can be neglected. If we assume $\Gamma_{2}=0$, Eq.\ (\ref{eq:master}) is simplified as
\begin{equation}
\frac{dP_{N}}{dt}=RP_{N-1}+\Gamma_{1}(N+1)P_{N+1}-(R+\Gamma_{1}N)P_{N}\;.
\label{eq:one-atom-master}
\end{equation}
This equation corresponds to the well-known birth-death model \cite{birth-death-model}. In the steady state, we have $dP_{N}/dt=0$ for all $N$ and the solution is a Poisson distribution given by
\begin{equation}
P_{N}=\frac{1}{N!}e^{-\bar{N}}\bar{N}^{N}
\label{eq:one-atom-distribution}
\end{equation}
for all integers $N \geq 0$ with $\bar N=R/\Gamma_1$, the mean atom number. 

A rate equation for the {\em ensemble-averaged} atom number $\langle N\rangle$ in the trap can be derived from the Eq.\ (\ref{eq:one-atom-master}). Multiplication of $N$ to the both sides of Eq.\ (\ref{eq:one-atom-master}) followed by summation over $N$ gives a differential rate equation for $\langle N\rangle$: 
\begin{equation}
\frac{d}{dt}\langle N\rangle=R-\Gamma_{1}\langle N\rangle\;,
\label{eq:one-rate-eq}
\end{equation}
where the ensemble average is formally defined as
\begin{equation}
\langle f(N)\rangle \equiv \sum_N^\infty P_N f(N)\;.
\end{equation}
The formal solution of Eq.\ (\ref{eq:one-rate-eq}) in terms of an initial atom number $N_0$ is given by
\begin{equation}
\langle
N(t)\rangle=\frac{R}{\Gamma_{1}}+\left(N_0-\frac{R}{\Gamma_1}\right)e^{-\Gamma_{1}t}\;.
\label{eq:one-solution}
\end{equation}
In the steady state, $d\langle N \rangle/dt=0$ and thus we get $\langle N\rangle=\bar{N}=R/\Gamma_{1}$
from Eq.\ (\ref{eq:one-rate-eq}). 
Here the upper bar indicates a time average in the steady state.
From the properties of the Poisson distribution we get
the following relation between the variance and the mean.
\begin {equation}
\sigma^{2}=\bar{N}=\frac{R}{\Gamma_{1}}\;.
\label{eq:one-relation}
\end {equation}

In the steady state, the correlation function of the atom number is defined as follows.
\begin{equation}
C(\tau)\equiv \langle N(t)N(t+\tau)\rangle_{t}
\label{eq:cor-def}
\end{equation}
where $\tau$ is a time delay and the notation $\langle \ldots \rangle_{t}$ represents a time average. Although the ensemble-averaged atom number $\langle N\rangle$ does not change in the steady state, equal to the mean atom number $\bar N$, the atom number $N$ itself is continuously fluctuating around its mean. 

We can replace the time average above with an ensemble average. Let us denote the atom number at time $t$ in the $i$th ensemble member as $N^{(i)}(t)$. Then we can rewrite the correlation function as
\begin{equation}
C(\tau)=\sum_i N^{(i)}(0)N^{(i)}(\tau)\;,
\end{equation}
where the summation over $i$ is performed over all members of the ensemble. We can regroup the ensemble into sub-ensembles in such a way that the members in each sub-ensemble have a common initial atom number $N_0$. Since $N^{(i)}$ values are distributed according to $P_N$ of Eq.\ (3), we can rewrite the above equation as
\begin{equation}
C(\tau)=\sum_{N_{0}}N_0 P_{N_{0}}\left[\sum_{i'} \left. N^{(i')}(\tau)\right|_{N^{(i')}(0)=N_0}\right]\;,
\label{eq:cor-def2}
\end{equation}
where the summation over $i'$ represents a summation over each sub-ensemble. All sub-ensembles are statistically identical, described by the same master equation, Eq.\ (2). Now, we can see that the quantity in $[\ldots]$ is nothing but the ensemble-averaged atom number $\langle N\rangle$ at time $t=\tau$ when its initial value is $N_0$ at time $t=0$.
The correlation function is then simplified as
\begin{equation}
C(\tau)=\sum_{N_{0}}^{\infty}N_{0}P_{N_{0}}\left.\langle N(\tau)\rangle \right|_{\langle N(0)\rangle=N_0}\;.
\label{eq:cor-def3}
\end{equation}


A differential equation for the correlation function can be obtained by taking a derivative of Eq.\ (\ref{eq:cor-def3}) with respect to $\tau$.
\begin{eqnarray}
\frac{d C(\tau)}{d
\tau}&=&\sum_{N_{0}}^{\infty}N_{0}P_{N_{0}}\frac{d}{d\tau}
\left.\langle N(\tau)\rangle \right|_{\langle N(0)\rangle=N_0}
\nonumber\\
&=&\sum_{N_{0}}^{\infty}N_{0}P_{N_{0}}\left\{R-\Gamma_{1}
\left.\langle N(\tau)\rangle \right|_{\langle N(0)\rangle=N_0}
\right\}\nonumber\\
&=&R\bar{N}-\Gamma_{1}C(\tau) \label{eq:one-cor-diff-eq}
\end{eqnarray}
In the second line in Eq.\ (\ref{eq:one-cor-diff-eq}), we used Eq.\ (\ref{eq:one-rate-eq}), which is independent of the initial condition. By using the conditions $C(0)=\langle N^2\rangle_t\equiv\overline{N^{2}}$ and $\bar{N}=R/\Gamma_{1}$, we obtain
\begin{equation}
\left.\frac{dC(\tau)}{d\tau}\right|_{\tau=0}=R\bar N - \Gamma_1  \overline{N^2} =R\bar N -\Gamma_1 ({\bar N}^2 +\bar N)=-R
\label{eq:one-cor-slope}
\end{equation}
and
\begin{equation}
C(\tau)={\bar{N}}^{2}+\sigma ^{2}e^{-\Gamma_{1}\tau}\;,
\label{eq:one-cor-sol}
\end{equation}
and therefore, the normalized correlation function is
\begin{equation}
C(\tau)\equiv\frac{\langle N(t)N(t+\tau)\rangle}{\langle
N(t)\rangle^{2}}=1+\frac{1}{\bar N}e^{-\Gamma_{1}\tau}.
\label{eq:one-nor-cor}
\end{equation}

The correlation function shows a decay behavior from $1+1/\bar N$ to 1 with a characteristic correlation time $t_{c}=1/\Gamma_{1}$. It is interesting to note that the atom-number correlation function exhibits bunching, {\em i.e.}, $C(0)>C(\tau)$ for $\tau>0$, when the atom number distribution is Poissonian whereas a Poisson distribution for photon number does not necessarily means bunching for a photon number correlation function. 

\section{With Two-atom Loss Terms}
If the density of atoms in a trap becomes higher, the two-atom loss terms are no longer negligible, and thus the full version of the master equation, Eq.\ (\ref{eq:master}), should be considered. However, the full master equation cannot be solved analytically. Instead, we rely on numerical solutions, either by Monte-Carlo simulation or an iteration method for given parameter values. Such numerical studies are discussed elsewhere \cite{sungsam-paper}. In this work, we focus on approximate solutions to the master equation.

The rate equation for $\langle N\rangle$ with the two-atom loss terms can be derived from the master equation as
\begin{equation}
\frac{d}{dt}\langle N\rangle=R-\Gamma_{1}\langle N\rangle-\Gamma_{2}\langle N(N-1)\rangle
\label{eq:two-rate-equation}
\end{equation}
In the steady state, we let $d \langle N \rangle /dt=0$. Differently from the case of excluding two-atom loss terms, however, we do not get any information about $\bar{N}$ other than the following relation between the variance and the mean.
\begin{equation}
\sigma^{2}=\frac{R}{\Gamma_{2}}+\left(1-\frac{\Gamma_{1}}{\Gamma_{2}}\right)\bar{N}-\bar{N}^{2}
\label{eq:two-relation}
\end{equation}



Following the same line of reasoning from Eq.\ (\ref{eq:cor-def}) to Eq.\ (\ref{eq:one-cor-diff-eq}) with the rate equation Eq.\
(\ref{eq:two-rate-equation}), we can obtain a differential equation for the correlation function in the presence of non-negligible two-atom loss terms as follows.
\begin{equation}
\frac{dC(\tau)}{d\tau}=R\bar{N}+(\Gamma_{2}-\Gamma_{1})C(\tau)-\Gamma_{2}\langle
N(t)N^{2}(t+\tau)\rangle
\label{eq:two-cor-diff-eq}
\end{equation}
When $\tau=0$, the slope of the correlation function becomes
\begin{equation}
\left.\frac{dC(\tau)}{d\tau}\right|_{\tau=0}=-R\bar N+(\Gamma_2-\Gamma_1)\overline{N^2}-\Gamma_2\overline{N^3}\;,
\label{eq:two-cor-slope}
\end{equation}
where $\overline{N^3}=\langle N^3\rangle_t$. The relation of $\overline{N^3}$ with the lower moments is obtained from the rate equation for $\langle N^2\rangle$ in the steady state.
\begin{equation}
\frac{d}{dt}\langle N^2\rangle=0=R(2\bar N+1)-\Gamma_{1}(2 \overline{N^2}-\bar N)-2\Gamma_{2}(\overline{N^3}-2\overline{N^2}+\bar N) \;.
\label{eq:two-rate-equation2}
\end{equation}
After some lengthy algebra we finally obtain
\begin{equation}
\left.\frac{dC(\tau)}{d\tau}\right|_{\tau=0}=-\frac{3}{2}R+\frac{\Gamma_{1}}{2}\bar{N}.
\label{eq:two-cor-slope}
\end{equation}
If $\Gamma_{2}=0$, Eq.\ (\ref{eq:two-cor-slope}) reduces to Eq.\ (\ref{eq:one-cor-slope}).

The differential equation Eq.\ (\ref{eq:two-cor-diff-eq}) cannot be solved exactly because of the complexity in the last term. However, in the limit of reasonably small two-atom loss, $\Gamma_{2} \ll R, \Gamma_{1}$, the atom number distribution is approximately a Poisson distribution with a variance $\sigma^2\approx \bar{N}$. This is the most significant approximation in our analysis. The validity of this approximation will be discussed in the next section.
In this limit, the mean value $\bar{N}$ can be obtained from Eq.\ (\ref{eq:two-relation}).
\begin{equation}
\bar N \approx \frac{R}{\Gamma_{2}}+\left(1-\frac{\Gamma_{1}}{\Gamma_{2}}\right)\bar{N}-\bar{N}^{2}
\end{equation}
and thus
\begin{equation}
\bar{N}\approx \frac{-\Gamma_1+\sqrt{\Gamma_1^2+4R\Gamma_2}}{2\Gamma_2}=\frac{2R}{\Gamma_{1}+\Gamma'}\equiv N'\;,
\label{eq:two-mean}
\end{equation}
where $\Gamma'=\sqrt{\Gamma_{1}^{2}+4R\Gamma_{2}}$. Equation (\ref{eq:two-mean}) reduces to Eq.\ (7) when $\Gamma_2\rightarrow 0$.

Our second approximation is that the correlation function is in the same form as that of Eq.\ (\ref{eq:one-cor-sol}) except for the decay constant. This is a reasonable approximation since the correlation function always starts from $\overline{N^{2}}$ and goes to $\bar{N}^{2}$ at infinity. 
Therefore, we assume
\begin{equation}
C(\tau)\simeq {\bar{N}}^{2}+\sigma^{2}e^{-\Gamma_{\rm eff}\tau},
\label{eq:two-cor-approx}
\end{equation}
where $\Gamma_{\rm eff}$ is an effective decay constant to be determined. It can be determined by Eq.\ (\ref{eq:two-cor-slope}).
\begin{equation}
-\sigma^2 \Gamma_{\rm eff} \simeq -\bar N \Gamma_{\rm eff} \simeq -\frac{3}{2}R+\frac{\Gamma_1}{2}\bar N
\label{eq:gamma-eff}
\end{equation}
and thus
\begin{equation}
\Gamma_{\rm eff}\simeq \frac{3R}{2\bar N}-\frac{\Gamma_1}{2}=\frac{\Gamma_{1}+3\Gamma'}{4}.
\label{eq:gamma-eff}
\end{equation}
The normalized correlation function is then given by
\begin{equation}
C(\tau)\simeq 1+\frac{\sigma^2}{\bar{N}^2}e^{-(\Gamma_{1}+3\Gamma')\tau/4}\;.
\label{eq:two-nor-cor}
\end{equation}
Using the approximation $\sigma^2\approx \bar N$, we obtain
\begin{equation}
C(\tau)\approx 1+\frac{1}{N'}e^{-(\Gamma_{1}+3\Gamma')\tau/4}\;.
\label{eq:two-nor-cor2}
\end{equation}

\begin{figure}
\includegraphics[width=3.3in]{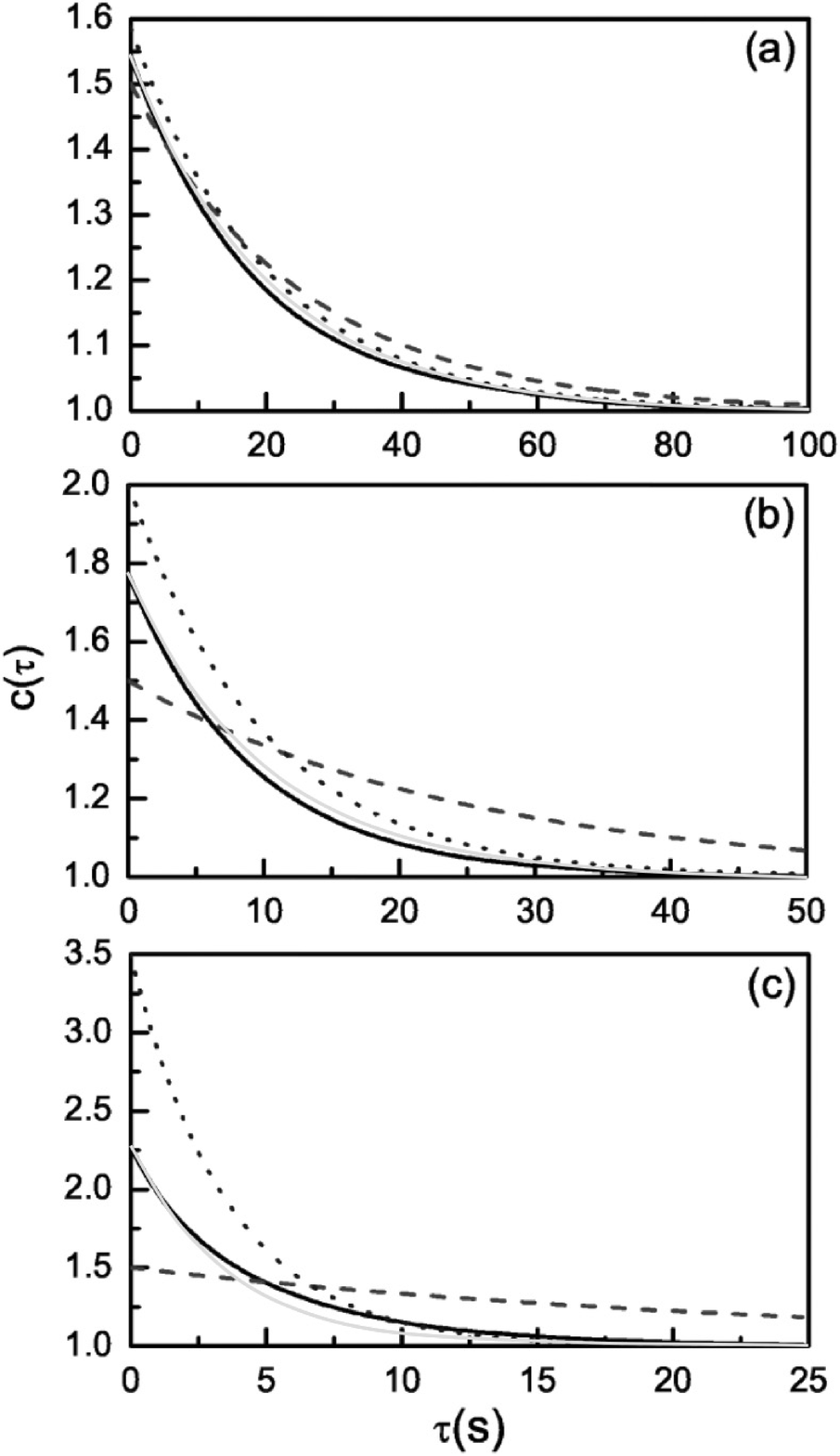}
\caption{Comparison of atom-number correlation functions by numerical simulation (solid curve) and by approximate formula (dotted curve).
Parameters are $R$=0.08, $\Gamma_{1}$=0.04 and the two-atom collision constant $\Gamma_{2}$ is (a) 0.004, (b) 0.04 and (c) 0.4, respectively. A dashed curve represents a correlation function given by Eq.\ (\ref{eq:one-nor-cor}) neglecting two-atom loss. Grey curves are the fit given by the approximate formula with $\sigma^2$ treated as a fitting parameter. The resulting $\sigma^2$ values are 0.94$\bar N$, 0.78$\bar N$ and 0.50$\bar N$ for (a), (b) and (c), respectively.}
\label{fig:compare}
\end{figure}

\section{Comparison with Simulation Results}
In order to check the validity of our approximate formulas Eqs.\ (\ref{eq:two-mean}) and (\ref{eq:two-nor-cor2}), we solve the master equation numerically and compare the results with the approximate formulas in Fig.\ \ref{fig:compare}. 
Under the condition of $\Gamma_{2} \ll R, \Gamma_{1}$, Eq.\ (\ref{eq:two-nor-cor2}) agrees well with the numerical results 
as shown in Fig.\ \ref{fig:compare}(a). Our approximation is valid in this limit. As $\Gamma_{2}$ is increased beyond this limit, however, the approximation starts to deviate from the simulation result as seen in Fig.\ \ref{fig:compare}(b), where we have $\Gamma_{2} \sim R, \Gamma_{1}$. The approximation is still acceptable even under this condition although not as good as in Fig. \ref{fig:compare}(a). The deviation is severe in Fig. \ref{fig:compare}(c), under the condition of $\Gamma_{2} \gg R, \Gamma_{1}$. 

The failure of the approximation is mostly due to the fact that the atom number distribution is assumed to be a Poissonian. The atom
number distribution function is not well approximated by a Poisson distribution when $\Gamma_{2}$ becomes comparable to and larger
than $R$ and $\Gamma_1$, so the assumption, $\bar{N}\approx\sigma^{2}$, fails. In general, the variance $\sigma^{2}$ is smaller than the
mean value $\bar{N}$. Consequently, when $\tau=0$, the approximate result is always larger than the true value.
\begin{equation}
C(0)|_{\rm approx}=1+\frac{1}{\bar{N}}>1+\frac{\sigma^{2}}{\bar{N}^{2}}=C(0)
\label{eq:c(0)}
\end{equation}
If we treat $\sigma^2$ in Eq.\ (\ref{eq:two-nor-cor}) as a fitting parameter, we obtain better agreement between the numerical results and the approximate formula even for $\Gamma_2\gtrsim \Gamma_1, R $. For example, the grey curves in Fig.\ \ref{fig:compare} are the fit given by Eq.\ (\ref{eq:two-nor-cor}), agreeing well with the numerical results.  

\section{CONCLUSIONS}
We have derived approximate formulas for the mean atom number and the atom-number correlation function in the limit of reasonably small two-atom collision compared to the one-atom collision and the loading rates. The validity of the approximate formulas was confirmed by comparing them with numerical solutions.

This work was supported by National Research Laboratory Grant and by Korea Research Foundation Grants (KRF-2005-070-C00058).

\end{document}